\documentclass[twocolumn,prb,showpacs,superscriptaddress,preprintnumbers,amsmath,amssymb]{revtex4}

\usepackage{graphicx}
\usepackage{dcolumn}
\usepackage{bm}

\begin{document}


\title{Change of Antiferromagnetic Structure near a Quantum Critical Point in CeRh$_{\bm{1-x}}$Co$_{\bm{x}}$In$_{\bm{5}}$}

\author{M.~Yokoyama}
\author{N.~Oyama}
\affiliation{Faculty of Science, Ibaraki University, Mito 310-8512, Japan}
\author{H.~Amitsuka}
\author{S.~Oinuma}
\author{I.~Kawasaki}
\affiliation{Graduate School of Science, Hokkaido University, Sapporo 060-0810, Japan}
\author{K.~Tenya}
\affiliation{Faculty of Education, Shinshu University, Nagano 380-8544, Japan}
\author{M.~Matsuura}
\affiliation{Neutron Science Laboratory, Institute for Solid State Physics, The University of Tokyo, Tokai 319-1106, Japan}
\author{K.~Hirota}
\affiliation{Neutron Science Laboratory, Institute for Solid State Physics, The University of Tokyo, Tokai 319-1106, Japan}
\affiliation{Graduate School of Science, Osaka University, Toyonaka 560-0043, Japan}
\author{T.J.~Sato}
\affiliation{Neutron Science Laboratory, Institute for Solid State Physics, The University of Tokyo, Tokai 319-1106, Japan}

\date{\today}
             
\begin{abstract}
The elastic neutron scattering experiments were carried out on the solid solutions CeRh$_{1-x}$Co$_x$In$_5$ to clarify the nature of the antiferromagnetic (AF) state in the vicinity of the quantum critical point (QCP): $x_c \sim 0.8$. The incommensurate AF order with the wave vector of $q_h=(1/2,1/2,\sim 0.3)$ observed in pure CeRhIn$_5$ is weakly suppressed upon doping with Co, and a commensurate $q_c=(1/2,1/2,1/2)$ and an incommensurate $q_1=(1/2,1/2,\sim 0.42)$ AF structures evolve at intermediate Co concentrations. These AF orders are enhanced at $x=0.7$, and furthermore the $q_h$ AF order vanishes. These results suggest that the AF correlations with the $q_c$ and $q_1$ modulations are significantly enhanced in the intermediate $x$ range, and may be connected with the evolution of the superconductivity observed above $x\sim 0.3$.
\end{abstract}

\maketitle

\section{Introduction}
The interplay between magnetism and superconductivity (SC) in the vicinity of a quantum critical point (QCP) is one of the most intriguing subjects in the heavy-fermion systems.  The QCP is characterized by the phase transition at zero temperature, which is generated by suppressing a magnetically ordered state via changing pressure, magnetic field and chemical composition. The SC often emerges near the QCP, and it is therefore believed that the magnetic fluctuation enhanced near the QCP plays a crucial role in the formation of the Cooper pairs. The recently discovered heavy-fermion compounds CeMIn$_5$ (${\rm M=Rh}$ and Co; HoCoGa$_5$-type tetragonal structure) provide a unique opportunity to investigate this subject. CeCoIn$_5$ is an unconventional superconductor characterized by an anomalously large specific-heat jump ($\Delta C/\gamma T_c=4.3$) at  transition temperature $T_c=2.3\ {\rm K}$.\cite{rf:Petrovic2001} A strong Pauli-limiting effect gives rise to a first-order transition at SC critical field $H_{\rm c2}$ below 0.7 K.\cite{rf:Izawa2001,rf:Tayama2002} In addition, above $H_{\rm c2}$ the non-Fermi-liquid behavior is observed in the temperature variations of the specific heat and the resistivity, which is considered to be due to the effect of quantum-critical fluctuation induced in the proximity of antiferromagnetism.\cite{rf:Bianchi2003} On the other hand, CeRhIn$_5$ shows an incommensurate (IC) antiferromagnetic (AF) order below $T_N=3.8\ {\rm K}$ at ambient pressure, whose structure has been proposed to be a helical (spiral) with a propagation vector of $q_h=(1/2,1/2,0.297)$.\cite{rf:Hegger2000,rf:Bao2000} As pressure $P$ is applied, $T_N$ initially increases to 4 K ($P=0.8\ {\rm GPa}$), and then rapidly decreases with further increasing $P$. At the same time, the SC develops above $1-1.5\ {\rm GPa}$. $T_c$ increases with increasing $P$ and merges with $T_N$ at the critical pressure $P_c=1.9\ {\rm GPa}$. \cite{rf:Mito2001,rf:Knebel2006,rf:Chen2006,rf:Yashima2007}

It is interesting to investigate the relationship between the IC-AF and SC phases for CeRhIn$_5$ and CeCoIn$_5$. The specific-heat, magnetization, and resistivity measurements \cite{rf:Zapf2001,rf:Jeffries2005} for mixed compounds CeRh$_{1-x}$Co$_x$In$_5$ revealed that the IC-AF phase for pure CeRhIn$_5$ is weakly suppressed upon doping with Co, and then vanishes at the QCP: $x_c\sim 0.8$. The SC phase appears above $x= 0.4$, suggesting the coexistence of both phases for intermediate Co concentrations. The obtained $x-T$ phase diagram is quite similar to the $P-T$ phase diagram for pure CeRhIn$_5$. Recently, we found  from the neutron scattering experiments that a commensurate (C) AF order with a modulation of $q_c=(1/2,1/2,1/2)$ evolves in the IC-AF phase at $x\sim 0.4$.\cite{rf:Yoko2006,rf:Yoko2008} This suggests that the C-AF correlation is tightly coupled with the SC. We expect that further insight on the role of the C- and IC-AF correlations to the SC is obtainable from both the bulk and microscopic investigations on a wide $x$ range. In addition, it is remarkable that the recently proposed $x-T$ phase diagram \cite{rf:Kawamura2007} differs from previous one,\cite{rf:Zapf2001} while it involves the C-AF phase in intermediate $x$ range. We have thus performed the specific heat and neutron scattering experiments on CeRh$_{1-x}$Co$_x$In$_5$ with entire $x$ range to clarify the characteristics of AF states near the QCP and find similarity or difference between them and the previously reported ones.

\section{Experimental Details}
Single crystals of CeRh$_{1-x}$Co$_x$In$_5$ were grown from an In flux with appropriate amounts of Ce, Rh and Co as starting materials. We prepared rod-type samples along the tetragonal $[1\bar{1}0]$ direction (typical size of $\sim 1.6 \times 1.6\times 15\ {\rm mm^3}$) for neutron scattering measurements by cutting out from the large ingots of the single crystal, and small pieces of single crystal for specific heat measurements. The electron probe microanalysis (EPMA) measurements for the sample used in the specific heat and neutron scattering experiments indicate that the estimated Rh/Co concentrations $x$ significantly deviate from the nominal (starting) values, which occurs frequently in the sample grown by the flux technique. Similar deviation of $x$ is also observed in the samples investigated previously:\cite{rf:Yoko2006} $x$ for the samples used in bulk and neutron scattering measurements were estimated to be 0.45 and 0.53, respectively, while the nominal $x$ was 0.4. In fact, a crucial problem on the reproducibility caused by the large distribution of $x$ was found in several samples. We will thus use the samples with homogeneous distribution of $x$ being achieved within a few percent error, and adopt the $x$ value estimated from the EPMA measurements for each sample throughout this article.

Specific heat $C_P$ was measured using a thermal relaxation method in a commercial $^3$He refrigerator (Oxford Instruments) from 0.5 K to 100 K. Elastic neutron scattering experiments for $x=0.05(1), 0.23(2), 0.43(2), 0.53(2)$ and 0.70(3) were performed on the triple-axis spectrometers GPTAS (4G) and PONTA (5G) located at the JRR-3M research reactor of the JAEA, Tokai. The neutron momentum $k=3.83\ {\rm \AA}^{-1}$ was chosen with the (002) reflection of the pyrolytic graphite (PG) for both monochromator and analyzer, and a set of 40'-40'-40'-80' collimators together with two PG filters was used. The samples were mounted in standard Al capsules filled with $^4$He gas so that the scattering plane becomes $(hhl)$, and cooled down to 1.5 K in a pumped $^4$He cryostat. In this geometry the sample rod is perpendicular to the scattering plane, enabling to minimize an effect of high neutron absorption by Rh and In. The neutron penetration length in the samples for the present $k$ value was calculated to be $\sim 1.8\ {\rm mm}$.

\section{Results and Discussion}
Figure 1 shows temperature variations of the specific heat divided by temperature $C_P/T$. A jump associated with the IC-AF order is clearly seen at $T_{Nh}=3.8\ {\rm K}$ for $x=0.05$. The magnitude of the jump is reduced with increasing $x$, and then the jump becomes obscure around $\sim 2.8\ {\rm K}$ at $x=0.7$. A kink anomaly appears at $T_{Nc}\sim 2.8\ {\rm K}$ in $C_P$ for $0.22 \le x \le 0.45$ (see the inset of Fig.\ 1), which originates from the C-AF order.\cite{rf:Yoko2006} The evolution of the SC yields another jump in $C_P/T$ at $T_c=1.4-2\ {\rm K}$ for $x \ge 0.29$, whose magnitude increases with increasing $x$. It seems that the entropy release estimated from $C_P/T$ below $T_c$ is not simply balanced. This indicates the existence of large residual entropy below $T_c$, which is actually evidenced in pure CeCoIn$_5$ from the $C_P/T$ data taken under strong magnetic fields.\cite{rf:Petrovic2001} This feature may be linked with the increase in $C_P/T$ above $T_{Nh}$ for the lower Co concentrations, suggesting that the magnetic critical fluctuation develops around $x_c$ as a consequence of the suppression of the AF order. These results on $C_P$ are basically consistent with the previous reports.\cite{rf:Zapf2001,rf:Jeffries2005,rf:Kawamura2007}
\begin{figure}[tbp]
\begin{center}
\includegraphics[keepaspectratio,width=0.45\textwidth]{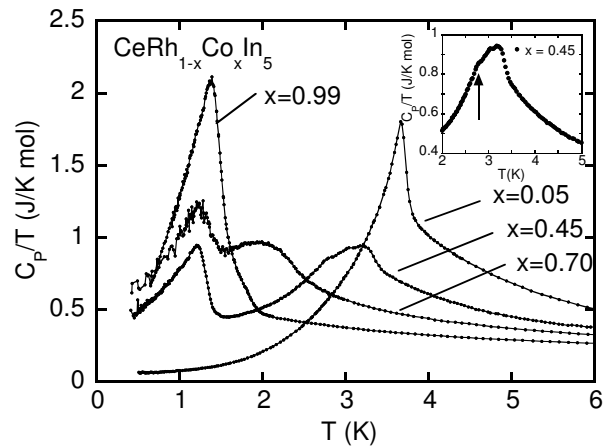}
\end{center}
  \caption{
Temperature variations of the specific heat divided by temperature $C_P/T$ for CeRh$_{1-x}$Co$_{x}$In$_5$. The enlargement around 3 K for $x=0.45$ is shown in the inset.
}
\end{figure}

In Fig.\ 2, we plot the $x-T$ phase diagram estimated from our $C_P/T$ data. Here we define $T_{Nh}$ and $T_{c}$ as the onset of the jump in $C_P/T$, and $T_{Nc}$ as the position of the kink anomaly. The $T_{Nh}$ and $T_c$ values as well as the shape of the $x-T$ phase diagram of ours are reasonably comparable to those in the first report.\cite{rf:Zapf2001} On the other hand, although an ambiguity on the determination of the transition temperatures still remains, there is a qualitative difference between our $x-T$ phase diagram and recently proposed one:\cite{rf:Kawamura2007} former involves both the IC- and C-AF phases in the intermediate $x$ range while latter does only the C-AF phase. The strong sample quality dependence may be responsible for such differences. Nevertheless, it is interesting that our $x-T$ phase diagram for CeRh$_{1-x}$Co$_x$In$_5$, including the existence of the C-AF order and the SC, is quite similar to that for the isostructural Ir-doped system CeRh$_{1-x}$Ir$_x$In$_5$.\cite{rf:Llobet2005} 
\begin{figure}[tbp]
\begin{center}
\includegraphics[keepaspectratio,width=0.45\textwidth]{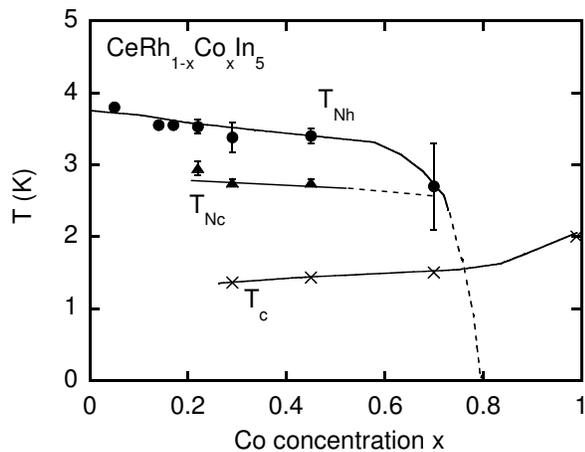}
\end{center}
  \caption{
The $x-T$ phase diagram estimated from the $C_P/T$ data for CeRh$_{1-x}$Co$_{x}$In$_5$. $T_{Nh}$ and $T_c$ are determined by the onset of the jump in $C_P/T$, while $T_{Nc}$ by the kink anomaly. The lines are guides to the eye.
}
\end{figure}

Displayed in Fig.\ 3 are neutron scattering patterns for the momentum transfers $Q=(1/2,1/2,1+\zeta)$ ($0\le \zeta \le 1$) obtained at 1.5 K, where the instrumental backgrounds were carefully subtracted using data at 5 K. A set of satellite Bragg peaks characterized by a single wave vector $q_h=(1/2,1/2,0.295(3))$ was observed for $x=0.05$. Similar peaks also appear at $x=0.23$, whose modulation corresponds to that for $x=0.05$ within the experimental accuracy. These peaks can be attributed to the occurrence of the IC-AF order, the same as that found in pure CeRhIn$_5$.\cite{rf:Bao2000} We cannot find any peak due to the other AF order at $x=0.23$, while $C_P$ at $x=0.22$ shows the kink at  $T_{Nc}$ as well as the jump at $T_{Nh}$. These Co concentrations are the boundary on the occurrence of the kink in $C_P$. We thus consider this inconsistency coming from a few percent error of $x$ in these samples. At $x=0.43$, we observed the evolution of new AF Bragg peaks ascribed to the AF order with the C $q_c=(1/2,1/2,1/2)$ structure together with the IC-AF Bragg peaks. The modulation of the IC-AF order slightly changes to $q_h=(1/2,1/2,0.303(4))$. In addition, weak peaks are detected at $Q=(1/2,1/2,0.40(1))$ ($\equiv q_1$) and its corresponding positions. However this order is considered to occur in a shorter range along the $c$ axis because the widths of these Bragg peaks are very large. These results for $x=0.43$ reproduce fairly well our previous report.\cite{rf:Yoko2006} We confirmed using polarized neutron scattering technique that all of these peaks originate from the alignments of the magnetic moments.\cite{rf:Yoko2008} At $x=0.7$, on the other hand, the AF orders with $q_c$ and $q_1=(1/2,1/2,0.416(5))$ modulations are enhanced, and simultaneously the $q_h$-AF order vanishes. The widths of these Bragg peaks seem to approach the resolution limit (about a few hundred {\rm \AA} in the length). We also checked a possibility of the other AF state appearing by the scan at $Q=(0,0,1+\zeta)$ ($0\le \zeta \le 1$) for $x=0.43$ and 0.7, but no additional Bragg peak is detected.
\begin{figure}[tbp]
\begin{center}
\includegraphics[keepaspectratio,width=0.45\textwidth]{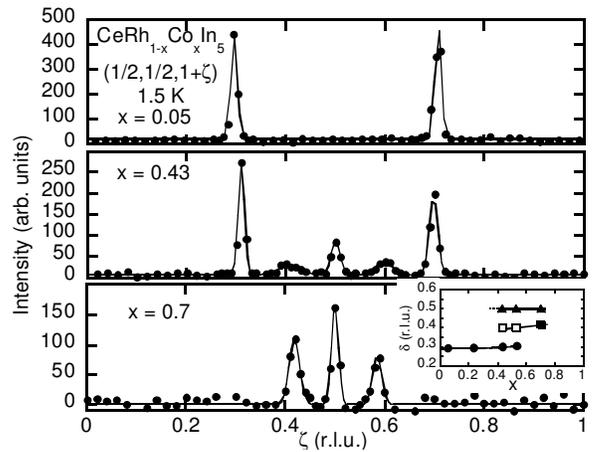}
\end{center}
  \caption{
Neutron scattering pattern at 1.5 K for CeRh$_{1-x}$Co$_x$In$_5$ with $x=0.05$, 0.43 and 0.7, obtained by scans at $Q=(1/2,1/2,1+\zeta)$ ($0\le \zeta \le 1$). The inset shows $x$ variations of the AF modulation $q=(1/2,1/2,\delta)$.
}
\end{figure}

The magnitudes of the volume-averaged AF moments $\mu_{h}$, $\mu_1$ and $\mu_{c}$ for the $q_h$, $q_1$ and $q_c$ structures, respectively, were estimated from the integrated intensities of the magnetic Bragg peaks obtained by the longitudinal and transverse scans, where the intensities of the weak nuclear (110) Bragg peak were used as a reference. In accordance with the previous reports,\cite{rf:Bao2000,rf:Yoko2006,rf:Majumdar2001} we assume that both the IC-AF orders have the helical structures, and the ordered moments in all the AF states lie in the tetragonal basal plane [Fig.\ 4(a)]. For all the AF states, the AF Bragg-peak intensities divided by the polarization factor calculated on the basis of above assumptions roughly follow the $|Q|$ dependence of the Ce$^{3+}$ magnetic form factor.\cite{rf:Brume62} The $x$ variations of $\mu_{h}$, $\mu_1$ and $\mu_{c}$ at 1.5 K are shown in Fig.\ 4(b). The $\mu_{h}$ values at $x=0.05$ and 0.23 are estimated to be 0.51(5) $\mu_{B}/{\rm Ce}$ and 0.59(2) $\mu_{B}/{\rm Ce}$, which are close to the values ($0.59-0.75\ \mu_{B}/{\rm Ce}$) at $x=0$.\cite{rf:Bao2000,rf:Raymond2007} For $x\ge 0.43$, $\mu_1$ and $\mu_{c}$ increase in connection with a reduction of $\mu_{h}$, and reach 0.27(2) $\mu_{B}/{\rm Ce}$ and 0.29(2) $\mu_{B}/{\rm Ce}$ at $x=0.7$, respectively. The total AF moment $\mu_{T}\ [\equiv (\mu_{h}^2+\mu_1^2+\mu_{c}^2)^{1/2}]$ shows a weak decrease as $x$ approaches $x_c\sim 0.8$.
\begin{figure}[tbp]
\begin{center}
\includegraphics[keepaspectratio,width=0.45\textwidth]{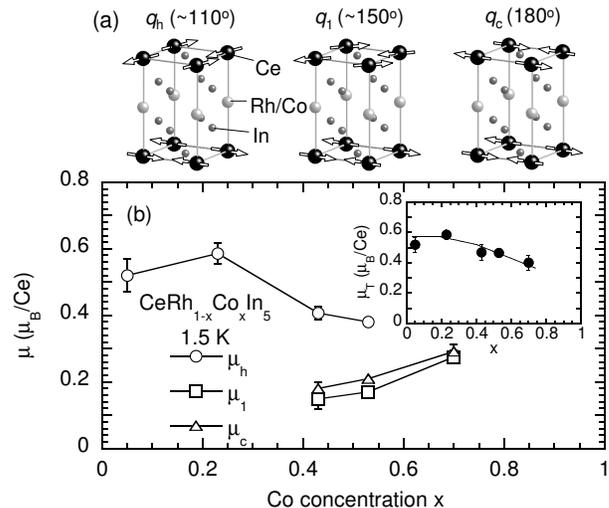}
\end{center}
  \caption{
(a) Arrangements of the magnetic moment on Ce ions for helical ($q_h$ and $q_1$) and commensurate ($q_c$) AF structures, and (b) $x$ variations of the AF moments $\mu_{h}$,  $\mu_{\rm 1}$ and  $\mu_{c}$ at 1.5 K for CeRh$_{1-x}$Co$_x$In$_5$. An angle between nearest neighbor moments along the $c$ axis for each structure is also shown in (a). The inset in (b) indicates $x$ variations of the total AF moment at 1.5 K, defined by $\mu_{T}=(\mu_{h}^2+\mu_{\rm 1}^2+\mu_{c}^2)^{1/2}$.
}
\end{figure} 

Temperature variations of the AF Bragg-peak intensities $I_{h}(T)$, $I_1(T)$ and $I_{c}(T)$ for $q_h$, $q_1$ and $q_c$ structures are plotted in Fig.\ 5. We observed that $I_{h}(T)$ for $x=0.05$ and 0.23 start increasing at 3.8 K, and then show a tendency to be constant below $\sim 3\ {\rm K}$.  At $x=0.43$, comparable $I_{h}(T)$ function is seen, but its onset is reduced to $3.4\ {\rm K}$. The Bragg-peak intensity due to the $q_{1}$-AF ordering (but short-ranged) coincidentally develops at almost the same temperature. The onsets of $I_{h}(T)$ agree with $T_{Nh}$ defined by the jump anomalies in $C_{P}$. $I_{c}(T)$ for $x=0.43$ starts increasing at 2.7 K, where we observed the kink anomaly in $C_{P}$. $I_{c}(T)$  follows a $T$-linear function in the wide temperature range. At $x=0.7$, on the other hand, the onset of $I_{ 1}(T)$ gets close to that of $I_{c}(T)$ around 2.9 K, where the broadened jump anomaly is observed in $C_P$. The different types of AF correlations exist with nearly the same energy scale, which may be responsible for such temperature dependence of $I_{c}(T)$. It should be noted that none of the $I_{h}(T)$ and $I_1(T)$ curves for all $x$ shows the reduction due to the appearance of the C-AF state. This strongly suggests that the C-AF order does not replace the IC-AF order but coexists with it in the sample, although it is still unclear whether the coexistence occurs in microscopic scale or not. 
\begin{figure}[tbp]
\begin{center}
\includegraphics[keepaspectratio,width=0.45\textwidth]{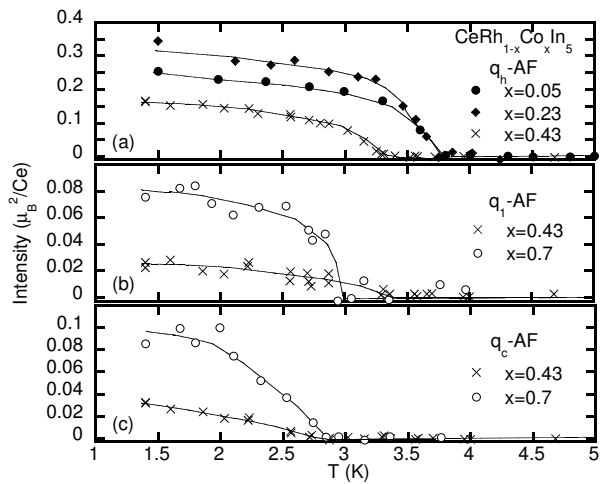}
\end{center}
  \caption{
Temperature variations of the AF Bragg-peak intensities for (a) $q_h$, (b) $q_1$ and (c) $q_c$ structures in CeRh$_{1-x}$Co$_x$In$_5$, which are mainly estimated at the $Q=(1/2,1/2,\sim 1.3)$, $(1/2,1/2,\sim 1.41)$, and (1/2,1/2,3/2) positions, respectively. The lines are guides to the eye.
}
\end{figure}   

It is observed that the AF orders with the $q_c$ and $q_1$ modulations simultaneously develop with increasing $x$, followed by the suppression of the $q_h$-AF order. This indicates that the AF correlations with the C modulation $q_c$ and its neighbor $q_1$ are significantly enhanced as $x$ approaches $x_c$: the quantum critical behavior is generated by the suppression of the AF order with $q_c$ and $q_1$ rather than the $q_h$-AF order, and may be connected with the evolution of the SC. Such a variation of the AF correlation is expected to influence strongly the low-energy magnetic fluctuations. For CeRhIn$_5$, it is pointed out that the AF fluctuation with the $q_h$ modulation develops at temperature much higher than $T_{Nh}$, and this may yield the increase in $C_P/T$ above $T_{Nh}$.\cite{rf:Bao2002} Recent inelastic neutron scattering experiments using a high energy-precision spectrometer revealed that the modulation of the AF fluctuation does not concentrate on $q_h$, but spreads widely along the $q=(1/2,1/2,\zeta)$ line in the $q$-space.\cite{rf:Aso2008} For CeCoIn$_5$, on the other hand,  recent inelastic neutron scattering experiments revealed the evolution of the magnetic excitation with the C $q_c$ modulation associated with the SC transition,\cite{rf:Stock2008} which is considered to be the evidence of the coupling between the C $q_c$-AF correlation and the SC. We suggest from these features on the AF fluctuation that the AF correlations with various propagation vector at around $\sim q_c$ on the $q=(1/2,1/2,\zeta)$ line cooperatively (or competitively) emerge in the intermediate $x$ range of CeRh$_{1-x}$Co$_x$In$_5$, leading to the complicated AF orders at $x \sim x_c$. Recent de Haas van Alphen experiments for CeRh$_{1-x}$Co$_x$In$_5$ revealed a dramatical change in the topology of the Fermi surface at much lower Co concentration than $x_c$,\cite{rf:Goh2008} which is expected to strongly influence the variations of the AF structure presently observed.

A simple relationship between the AF order and the SC rather than that seen in CeRh$_{1-x}$Co$_x$In$_5$ is realized in the CeCo(In,Cd)$_5$ system.\cite{rf:Pham2006} Doping Cd into CeCoIn$_5$ is found to suppress the SC phase, and then generate the AF phase above $\sim$ 10\% Cd concentration. This AF phase involves only the C $q_c$ structure.\cite{rf:Nicklas2007} In addition, the AF transition temperature monotonically increases to $\sim 5\ {\rm K}$ with increasing the Cd content up to 30 \%. These properties of the AF state contrast sharply with those found in CeRh$_{1-x}$Co$_x$In$_5$, where we observed the AF orders having both C and IC modulations and the AF transition temperature to be kept at $\sim 3.8\ {\rm K}$ as $x$ decreases from $x_c$. We consider that such differences may be due to the effect of the IC-AF correlations with $q_h$  involved in CeRh$_{1-x}$Co$_x$In$_5$.

\section{Summary}
The present neutron scattering and specific heat measurements for CeRh$_{1-x}$Co$_x$In$_5$ confirmed that both the AF and SC appear in a wide $x$ range, which is consistent with the previous reports.\cite{rf:Zapf2001,rf:Jeffries2005} The AF order with the $q_h$ structure is suppressed with increasing $x$, and both the $q_c$ and $q_1$ structures emerge in the intermediate $x$ range. This suggests that the AF correlations with the C $q_c$ and its neighbor $q_1$ may be tightly coupled with the evolution of the SC.

\section*{Acknowledgments}
We thank N.\ Aso for informative discussions. This work was supported by a Grant-in-Aid for Scientific Research from the Ministry of Education, Culture, Sports, Science and Technology of Japan.



\begin{thebibliography}{99}
\bibitem{rf:Petrovic2001}C.\ Petrovic, P.\ G.\ Pagliuso, M.\ F.\ Hundley, R.\ Movshovich, J.\ L.\ Sarrao, J.\ D.\ Thompson, Z.\ Fisk and P.\ Monthoux, J.\ Phys.: Condens.\ Matter {\bf 13}, L337 (2001).
\bibitem{rf:Izawa2001}K.\ Izawa, H. Yamaguchi, Y. Matsuda, H. Shishido, R. Settai and Y. Onuki, Phys. Rev. Lett. {\bf 87}, 057002 (2001).
\bibitem{rf:Tayama2002}T.\ Tayama, A. Harita, T. Sakakibara, Y. Haga, H. Shishido, R. Settai and Y. Onuki, Phys.\ Rev.\ B {\bf 65}, 180504(R) (2002). 
\bibitem{rf:Bianchi2003}A.\ Bianchi,  R.\ Movshovich, I.\ Vekhter, P.\ G.\ Pagliuso and J.\ L.\ Sarrao, Phys.\ Rev.\ Lett {\bf 91}, 257001 (2003).
\bibitem{rf:Hegger2000}H.\ Hegger, C.\ Petrovic, E.\ G.\ Moshopoulou, M.\ F.\ Hundley, J.\ L.\ Sarrao, Z.\ Fisk and J.\ D.\ Thompson, Phys.\ Rev.\ Lett.\ {\bf 84}, 4986 (2000).
\bibitem{rf:Bao2000}W.\ Bao, P.\ G.\ Pagliuso, J.\ L.\ Sarrao, J.\ D.\ Thompson, Z.\ Fisk, J.\ W.\ Lynn and R.\ W.\ Erwin, Phys.\ Rev.\ B {\bf 62}, R14621 (2000);  Phys.\ Rev.\ B {\bf 67}, 099903(E) (2003).
\bibitem{rf:Mito2001}T.\ Mito, S.\ Kawasaki, G.-q.\ Zheng, Y.\ Kawasaki, K.\ Ishida, Y.\ Kitaoka, D.\ Aoki, Y.\ Haga and Y.\ Onuki,  Phys. Rev. B {\bf 63}, 220507(R) (2001).
\bibitem{rf:Knebel2006}G.\ Knebel, D.\ Aoki, D.\ Braithwaite, B.\ Salce and J.\ Flouquet, Phys. Rev. B {\bf 74}, 020501(R) (2006).
\bibitem{rf:Chen2006}G.\ F.\ Chen, K.\ Matsubayashi, S.\ Ban, K.\ Deguchi and N.\ K.\ Sato, Phys.\ Rev.\ Lett.\ {\bf 97}, 017005 (2006).
\bibitem{rf:Yashima2007}M.\ Yashima, S.\ Kawasaki, H.\ Mukuda, Y.\ Kitaoka, H.\ Shishido, R.\ Settai and Y.\ Onuki, Phys.\ Rev.\ B {\bf 76}, 020509(R) (2007).
\bibitem{rf:Zapf2001}V.\ S.\ Zapf, E.\ J.\ Freeman, E.\ D.\ Bauer, J.\ Petricka, C.\ Sirvent, N.\ A.\ Frederick, R.\ P.\ Dickey and M.\ B.\ Maple, Phys.\ Rev.\ B {\bf 65}, 014506 (2001).
\bibitem{rf:Jeffries2005}J.\ R.\ Jeffries, N.\ A.\ Frederick, E.\ D.\ Bauer, H.\ Kimura, V.\ S.\ Zapf, K.-D.\ Hof, T.\ A.\ Sayles and M.\ B.\ Maple, Phys.\ Rev.\ B {\bf 72}, 024551 (2005).
\bibitem{rf:Yoko2006}M.\ Yokoyama, H.\ Amitsuka, K.\ Matsuda, A.\ Gawase, N.\ Oyama, I.\ Kawasaki, K.\ Tenya and H.\ Yoshizawa, J.\ Phys.\ Soc.\ Jpn. {\bf 75}, 103703 (2006).
\bibitem{rf:Yoko2008}M.\ Yokoyama, N.\ Oyama, H.\ Amitsuka, S.\ Oinuma, I.\ Kawasaki, K.\ Tenya, M.\ Matsuura and K.\ Hirota, Physica B {\bf 403}, 812 (2008).
\bibitem{rf:Kawamura2007}S.\ Ohira-Kawamura, H.\ Shishido, A.\ Yoshida, R.\ Okazaki, H.\ Kawano-Furukawa, T.\ Shibauchi, H.\ Harima and Y.\ Matsuda, Phys. Rev. B {\bf 76}, 132507 (2007).
\bibitem{rf:Llobet2005}A.\ Llobet, A.\ D.\ Christianson, W.\ Bao, J.\ S.\ Gardner, I.\ P.\ Swainson, J.\ W.\ Lynn, J.-M.\ Mignot, K.\ Prokes, P.\ G.\ Pagliuso, N.\ O.\ Moreno, J.\ L.\ Sarrao, J.\ D.\ Thompson and A.\ H.\ Lacerda, Phys.\ Rev.\ Lett.\ {\bf 95}, 217002 (2005). 
\bibitem{rf:Majumdar2001}S.\ Majumdar, G.\ Balakrishnan, M.\ R.\ Lees, D.\ McK.\ Paul and G. J. McIntyre, Phys.\ Rev.\ B {\bf 66}, 212502 (2002).
\bibitem{rf:Brume62}M.\ Blume, A.\ J.\ Freeman and R.\ E.\ Watson, J.\ Chem.\ Phys.\ {\bf 37}, 1245 (1962).
\bibitem{rf:Raymond2007}S.\ Raymond, E.\ Ressouche, G.\ Knebel, D.\ Aoki and J.\ Flouquet, J.\ Phys.: Condens. Matter {\bf 19}, 242204 (2007).
\bibitem{rf:Bao2002}W.\ Bao, G.\ Aeppli, J.\ W.\ Lynn, P.\ G.\ Pagliuso, J.\ L.\ Sarrao, M.\ F.\ Hundley, J.\ D.\ Thompson and Z.\ Fisk, Phys.\ Rev.\ B {\bf 65}, 100505(R) (2002).
\bibitem{rf:Aso2008}N.\ Aso, private communications. 
\bibitem{rf:Stock2008}C.\ Stock, C.\ Broholm, J.\ Hudis, H.\ J.\ Kang and C.\ Petrovic, Phys.\ Rev.\ Lett.\ {\bf 100}, 087001 (2008).
\bibitem{rf:Goh2008}Swee K.\ Goh, J.\ Paglione, M.\ Sutherland, E.\ C.\ T.\ O'Farrell, C.\ Bergemann, T.\ A.\ Sayles and M.\ B.\ Maple, cond-mat/0803.4424 (unpublished).
\bibitem{rf:Pham2006}L.\ D.\ Pham, T.\ Park, S.\ Maquilon, J.\ D.\ Thompson and Z.\ Fisk, Phys.\ Rev. Lett.\ {\bf 97}, 056404 (2006).
\bibitem{rf:Nicklas2007}M.\ Nicklas, O.\ Stockert, T.\ Park, K.\ Habicht, K.\ Kiefer, L.\ D.\ Pham, J.\ D.\ Thompson, Z.\ Fisk and F.\ Steglich, Phys.\ Rev.\ B {\bf 76}, 052401 (2007).
\end{thebibliography}
\end{document}